# Is radicalization reinforced by social media censorship?

Justin E. Lane,[1] Kevin McCaffree[2], & F. LeRon Shults[1]

## Abstract

Radicalized beliefs such as those tied to the QAnon conspiracy theory can lead some individuals and groups to engage in violent behaviour, as evidenced by the attack on the US Capitol on 6 January 2021. Understanding the mechanisms by which such beliefs are accepted, spread, and intensified is critical for any attempt to mitigate radicalization and avoid increased political polarization. This article presents an agent-based model of a social media network that enables investigation of the effects of censorship on the amount of dissenting information to which agents become exposed and the certainty of their radicalized views. The model explores two forms of censorship: 1) decentralized censorship—in which individuals can choose to break an online social network tie (unfriend or unfollow) with another individual who transmits conflicting beliefs and 2) centralized censorship—in which a single authority can ban an individual from the social media network for spreading a certain type of belief. This model suggests that both forms of censorship increase certainty in radicalized views by decreasing the amount of dissent to which an agent is exposed, but centralized "banning" of individuals has the strongest effect on radicalization.

## Political Polarization and Social Media

Political polarization in the United States began to increase in the 1960s and accelerated rapidly in the 1990s and 2000s [1]. This polarization appears to have begun among political elites, who used controversial shortcuts to ease the process of re-election (e.g., gerrymandering districts, aligning party politics with evangelical religious doctrine, de-regulating campaign finance). Elected political officials now operate within an environment of extreme ideological polarization, which has spread to the public electorate [2]. This polarized environment can also lead to a deep cynicism about the political system in general; if horrible people on the "other side" are allowed to work in Washington, then perhaps the whole system is nothing but a swamp of incompetent, or even evil, individuals. Such cynicism provides fertile cultural soil for the emergence and spread of right-wing and left-wing populism, which seems to be driven in part by a disdain for political institutions [3].

---

[1] ALAN Analytics, s.r.o, Bratislava, Slovak Republic; Center for Modeling Social Systems, Kristiansand Norway; Center for Mind and Culture, Boston, MA
[2] University of North Texas

At least 60% of the American public gets their news from social media sites "at least on occasion," with nearly 20% relying on such sources "often" [4]. Facebook, in particular, seems to be leading this trend, with 67% of American adults using the site and 44% getting their news while doing so [4]. This is especially important given that 74% of Facebook users check their site on a daily basis and 51% check multiple times a day [5]. Despite the popularity of Facebook, data indicate that people are increasingly getting their news from social media sites of various kinds. In fact, Reddit, Facebook, Twitter, Tumblr, Instagram, YouTube and Vine all show increasing rates of news-seeking among users [4].

Social media sites are thus quickly becoming central platforms for news distribution [6], indicating that such sites have the potential to alter the larger cultural milieu [7]. In a polarized and cynical political environment, individuals driven by a desire for "chaos" and disruption will be motivated to generate and spread empirically false, politically hostile, news rumors online [8,9]. Empirically false news stories with a negative emotional valence spread more quickly and are shared more widely compared to empirically true news stories [10]. Although "bots" appear to be a considerable source of false news headlines, human beings—especially those who are heavy news media consumers—are more likely to spread fake news compared to bots [10,11].

Simply being exposed to false news stories increases the likelihood that people will accept such stories as accurate, even if such a story is "flagged" as problematic by third-party fact checkers [12]. Adding to this problem is that fake political news headlines tend to be more novel (i.e., informationally distinct from other tweets), and thus tend to attract more attention. Indeed, false political news headlines are about 70% more likely to retweeted on Twitter [10]. To the extent that the public is increasingly exposed to false political news, it is possible that some members of the public are self-radicalizing by filling their social media feeds with empirically false or misleading news stories that verify their existing political biases while vilifying members of other parties.

## Echo Chambers and Radicalization

Heavy political news consumers on social media may be creating "echo chambers" by curating news feeds and "friends" online who share their political views, which can exacerbate political polarization. A large body of research on the "homophily bias" (a cross-cultural tendency to associate with those most similar to self) provides prima facie validity of the idea of echo chamber construction [13]. Such an idea is also supported by findings that suggest politically-biased individuals are more likely to believe political news headlines which confirm their existing biases [14], and that left-leaning individuals are more likely to watch left-biased news networks (e.g., CNN) while right-leaning individuals are more likely to watch right-biased networks (e.g., Fox News) [15].

Echo chamber construction may also be driven by a "false consensus" effect. Individuals generally tend to overestimate the degree to which other individuals agree with their personal opinions, thus (unwittingly) creating the perception of consensus where, in fact, none may exist. Research has shown that participating in ideologically homogeneous social networks can increase this false consensus effect. For example, the false consensus bias among neo-Nazis increases (even when controlling for level of extremism) with individuals' participation in online discussion forums [16]. Further research on neo-Nazis recruited from online forums found that participating in dissimilar real-world networks did not decrease false consensus effects, but being exposed to dissimilar news media did [17]. While this research was conducted in an online environment that is different from networks such as Facebook and Twitter, it is nevertheless relevant for our purposes because it suggests that offline social networks do not necessarily have effects on the perception of false consensus, while exposure to different types of news media can. Additionally, research on the false consensus effect has found that individuals are more motivated to defend their position when they perceive their cause to be under threat, and their perception of a false consensus might therefore motivate them to defend their position [18,19].

The echo chamber hypothesis is also supported by research indicating that political partisans tend both to select news media which confirms their political biases as well as associate disproportionately online with people who share similar partisan biases [20,21]. Research also shows that self-reported use of social media platforms (Facebook, Twitter, Instagram) increases the perception of national political polarization, with those using social media more often perceiving greater polarization [22]. In other words, some people may be psychologically motivated to cultivate politically homophilous social media sites, a motivation that may be amplified as the probability of their being exposed to polarizing content while online increases.

Another line of research suggests that echo chambers filled with false, politically-polarized news media are not as common as is typically supposed. Due in large part to the rapidity and breadth of information diffusion in online environments (as well as the relative diversity of friendship networks on social media), some research indicates that social media exposes people to a greater diversity of news stories than they may otherwise see in an offline environment [23]. This is true even for heavy consumers of online news media. Although the construction of echo chambers is a real phenomenon, exposure to diverse news stories still functions as a counter-current to such ideological isolation. It is also important to note that internet users who retweet false news headlines have significantly fewer followers, follow significantly fewer people and tend to be less active on Twitter in general [10]. Other work in this area indicates that the most extreme false news echo chambers tend to be created and occupied by a relatively small proportion of internet users [11,24–26].

Other reasons for optimism can be gleaned from research showing that exposing an individual to two sides of an argument can serve to decrease the false consensus bias [27], and that Facebook usage fosters the development of network diversity (i.e., bridging ties), especially among users with low self-esteem [28,29]. Our argument here is not that all, or even most, people using social media for news-related information exist in echo chambers. Rather, we suggest that the phenomenon of echo chamber construction exists, that is may plausibly be growing alongside offline partisan polarization and that it can have substantial effects on peoples' ideological rigidity and extremism.

## Mechanisms of Radicalization and Forms of Social Media Censorship

Despite the caveats just mentioned, it seems clear that personalized social media sites (e.g., Facebook, Twitter, Instagram) containing politically self-similar "friends," "followers," and polarized news feeds have the capacity to initiate or contribute to a process of self-radicalization.

As we have seen, there is evidence suggesting that participation in online social networks can increase exposure to false, negatively valenced political news headlines, can amplify false consensus beliefs, and can facilitate the construction of news media echo chambers. However, there is currently no strong evidence regarding the extent to which online censorship can affect an individual's certainty about their own beliefs.

There are at least six mechanisms by which censorship might cause an increase in the probability of forming extreme or radical beliefs. These mechanisms include identity uncertainty [30], differential association and social bonding [31–35], shared negative affect and identity fusion [36,37] group polarization [38], mutual verification contexts [39] and internalized labeling [40,41].

When an individual is censored from expressing themselves on a social media platform, this can be experienced as a form of identity non-verification, i.e., a feeling that they have been denied the possibility of expressing their sense of identity in the future. Depending on how salient the identity being expressed was to the individual, such identity non-verification may increase his or her identity uncertainty [30]. Identity uncertainty is a state wherein a person feels incapable of predicting or controlling future interactions and circumstances due to a basic lack of certainty in who they are, why they are who they are, or why they are being treated in a given way on the basis of who they are. Identity uncertainty is an emotionally unpleasant experience. Under some conditions of identity uncertainty, individuals may therefore seek to fortify their identity by enhancing its degree of cohesiveness and clarity (i.e., they will seek to increase the certainty of their self-concept).

To further fortify their existing identity, people may also choose to associate with others who have similarly been censored, or people who have not yet been censored but who share identities with those who have been censored. Due to network homophily biases [13], banned individuals may thus differentially associate with self-similar others, creating social bonds with people whose ideas are similarly "unacceptable" (or deemed unacceptable by a given social media company's rules of conduct). As Hogg [30] notes, radical ideologies can be very satisfying to those with uncertain identities primarily because extremist ideologies are zero-sum and exclusivist, and therefore conducive to a rigid sense of identity certainty.

Once individuals have been censored, perceived this censorship to be an instance of emotionally-aversive identity non-verification, and formed bonds with self-similar others, they may begin to ruminate about their negative experiences of exclusion and negatively valenced emotions (guilt, shame) associated with social exclusion. Research at the intersection of cognitive psychology and social psychology indicates that when individuals ruminate about shared dysphoric experiences, they can experience an increased motivation to cooperate with one another, as well experience an increased fusion of their personal identity with that of the group [36] (in this case, theoretically, the group of people also censored from social media, or who also hold ideas eligible for censorship). Once a group identity begins to form among censored individuals, Sunstein's [38] "law of group polarization" suggests that awareness of self-similar others who share experiences of censorship (or concerns about the possibility of censorship) will significantly embolden more moderate group members who will, in turn, look to even more extreme group archetypes as models of ideal group member behavior.

This recursive within-group polarization dynamic can lead to hostile extremism, especially insofar as it leads to the emergence of "mutual verification contexts" [39], i.e., situations within which people derive positive identity-affirming emotions amongst a small group of others with shared identities. As a group based in shared dysphoric experiences interacts over a period of time, positive emotions begin to accrue on the basis of expectations for identity verification from other group members, who themselves expect identity verification. Thus, censorship can produce extremism insofar as mutual verification contexts (formed on the basis of differential association amongst those sharing dysphoric experiences) help encode identity certainty, and insofar as within-group polarization dynamics accelerate the development of more and more extreme beliefs.

Finally, labeling theory in criminology [40], suggests that people will sometimes act in accordance with the ways in which they are labeled by authorities. This approach leverages the notion of a self-fulfilling prophecy (that how others are expected to act, or believe themselves to be expected to act, influences how they eventually do act) to explain criminal behavior and ideological extremism [41]. Once censored for being deviant/bad/dangerous (however defined), individuals may be motivated to absorb this label into their existing self-concept in an attempt to interpret the censorious sanctions levied against them. Not only can this internalization of a deviant label motivate association with others so-labeled, it can begin to generate a certainty that one is, in fact, deviant or dangerous. To the extent that a person believes themselves to be dangerous or deviant, they may act in a way that expresses actual deviance or dangerousness so as to maintain a sense of self-consistency and self-understanding; known as the "cognitive consistency principle" [42]. Notably, a key tenet of labeling theory is that this process of internalized labeling can occur even if an individual had not acted in a deviant or dangerous manner (i.e., in the case of individuals being censored arbitrarily or indiscriminately)—the simple act of being labeled by an authority can lead to a process of identity re-construction and subsequent deviant or dangerous behavior.

In the present paper, we focus on three forms of censorship, any one of which might serve to increase radicalism according to the above six mechanisms. Some social networks give users the ability to choose to stop seeing information from other individuals by "blocking," "unfollowing," or "unfriending" them—we refer to these mechanisms as *decentralized censorship*. Alternatively, some social media platforms censor users unilaterally by suspending or banning them due to behaviors that violate their formal rules of conduct. For example, spreading hate speech or particularly offensive videos can result in being banned by Facebook and Twitter. Other online social networks, such as youtube.com and medium.com, have also censored members of far-right political persuasions. We call these *centralized censorship* mechanisms because in this case it is third-party administrators, working in a centralized fashion on behalf of a social media company, who are enacting the censorship. It also the case that social networks such as Facebook and Twitter allow for both forms of censorship. When both mechanisms are in use, we call this *mixed censorship*.

The extent to which these forms of censorship have tangible effects on radicalization is an unexplored research topic. While decentralized, centralized and mixed forms have censorship can have a positive effect in that they shield individuals—particularly targets of hate speech—from being confronted with offensive speech in the online community, they do not stop the individuals expressing such ideas from holding hateful beliefs. As such, it is conceivable that censoring individuals who share radical content only serves to cut off their opportunity to encounter dissenting viewpoints, which allows their radicalism to carry on unchecked (according to the six mechanisms discussed above). The online safety of users is clearly an important issue to consider and censorship appears to provide a means to stop offensive information from reaching those who find it offensive. However, we know very little about the wider social effects various forms of censorship might have.

## Model

In order to investigate the ways in which radicalized belief can be affected by different forms of censorship within online social networks, we constructed an agent-based model that can implement centralized, decentralized, and mixed censorship mechanisms and simulate their effects. The model was programmed in NetLogo [43]. First, agents (n = 100) were initialized onto a small world network to simulate earlier findings on the structure of human social networks [44–47].

We employed a unidirectional network to represent the fact that information flows on many social networks are not bi-directional (e.g., individuals who follow an account on Twitter are not required to be followed by that individual). Agents in the model are initialized with an initial belief (0 or 1), with belief "1" representing a radical belief. If a simulation run is utilizing mixed or centralized forms of authority to censor beliefs (AUTHORITY_CENSORSHIP = "Centralized" or "Mixed"), the model chooses one agent with a belief of "0" to act as the central authority. Agents interact with one another at each time step. Agents broadcast their belief to their social network ties. If the belief an agent receives is different from their own belief, they increase their level of dissent by 1 for each dissonant belief from their own. If the belief an agent receives is the same as their own belief, they increase their level of assent by 1 for each consonant belief of their own.

Agents then have a chance to create a new link with another agent in the network. If the model is running with centralized censorship mechanisms, and an agent has been banned by the central authority, they will create a link with another banned agent with a probability equal to the model's HOMOPHILY variable. This mechanism uses a random float number between 0 and 1. If the number is less than the HOMOPHILY variable, the condition for creating a link with a similar agent is considered satisfied. If the agent has not been banned, it will create a link to an agent with a similar belief with a probability equal to the model's HOMOPHILY variable. However, in this case, if the random float number is greater than HOMOPHILY, the agent will still make a link to an agent, but the link will be made with a random agent.

After the agents interact and create new links, agents can then be censored. There are two censorship mechanisms in this model: centralized and decentralised. If the variable AUTHORITYCENSORSHIP is set to "true," the model will run with centralized censorship. Otherwise, the model utilizes decentralized censorship. The centralized censorship mechanism utilizes a single "authorized" agent that can police any other agent in the model. At each time step the authorized agent will choose one other agent that has not yet been banned. If its belief is equal to 1 (the censored belief), the agent is banned and all their links to other agents are removed with a probability equal to the TOLERANCE variable. If the model is operating with decentralized censorship, all agents with belief 0 will access the belief of one of their out-links. If the belief of the selected out-linked agent is equal to 1, they will remove the link between themselves and that agent (representing unfriending on Facebook or unfollowing on Twitter). From the model, the average assent and dissent for each group (defined as holding the target belief, or not) is recorded, as is the average degree of network ties (the number of links that agent has to other agents in the network) of agents holding those respective beliefs. For ease of analysis, the divergence between the amount of assent and dissent for each belief is also reported. In addition, the agent's level of social certainty is calculated as the percentage of all links connected to the agent (regardless of direction) that share their belief.

## Analysis

The model was analyzed using a Latin Hypercube Sampling technique that drew random conditions from the possibility space of the model. From this output data, we investigated what general patterns could be discerned for each censorship style.

## Results

The first step in our analysis was to produce a correlation heatmap that could help us explore the effects of different variables on one another throughout the state space of the simulated experiment. The results of this are shown in Figure 1 below.

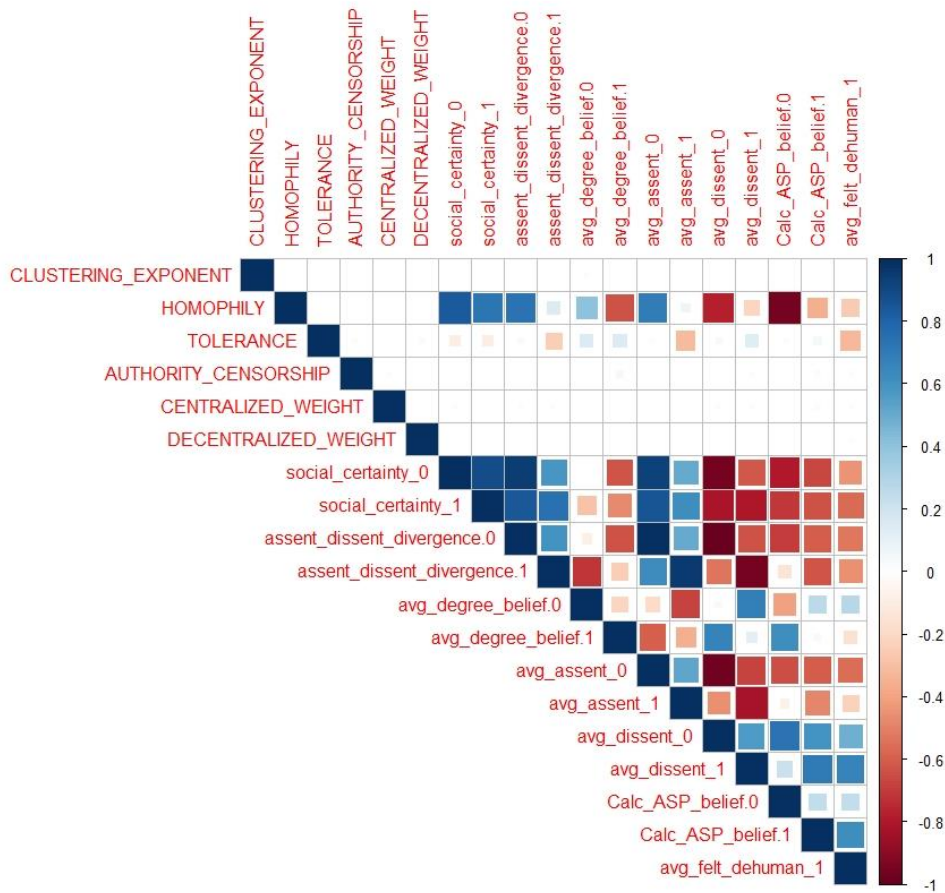

Figure 1. Correlation heatmap

A cursory review of the heatmap shows that homophily and tolerance (the thresholds for creating and dropping links in the network based on shared beliefs, respectively) appear correlated with effects that suggest the isolation of groups.

In addition to identifying correlations among variables, we utilized visualization techniques to help us explore the extent to which different censorship mechanisms had significant impacts on the way in which individuals formed groups and became more certain of their beliefs. The graphs presented in Figure 2 and 3 below suggest that the centralized and mixed censorship conditions largely behave in the same way. In non-radical groups (Figure 2), these forms of censorship resulted in higher certainty of agents' beliefs, while decentralized censorship resulted in greater belief variety.

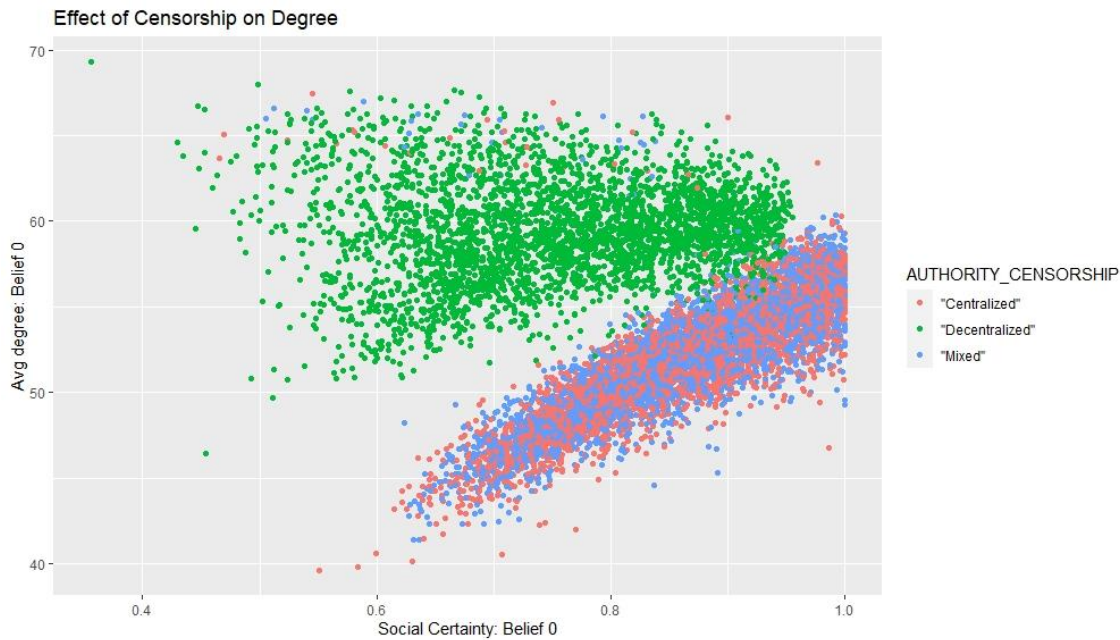

Figure 2. Effects of censorship on non-radicalized groups (belief 0 is the non-radicalized belief).

In groups with radicalized beliefs (Figure 3), the mixed and centralized censorship mechanisms behave in largely the same way as they do for non-radicalized groups, but a very different pattern emerged for decentralized censorship. Again, decentralized censorship revealed a wider variety of degree of network ties and certainty. However, the centralized and mixed censorship mechanisms tended to create a higher level of certainty in the agents. This was particularly the case when the average degree of network ties decreases (the number of average links an agent has), likely the result of the individual being censored on the platform (and links being dropped, causing a lowering of their degree). Visually depicting the results, as in Figure 2 and 3, demonstrates how these different censorship styles appear to have two very different clusters of results.

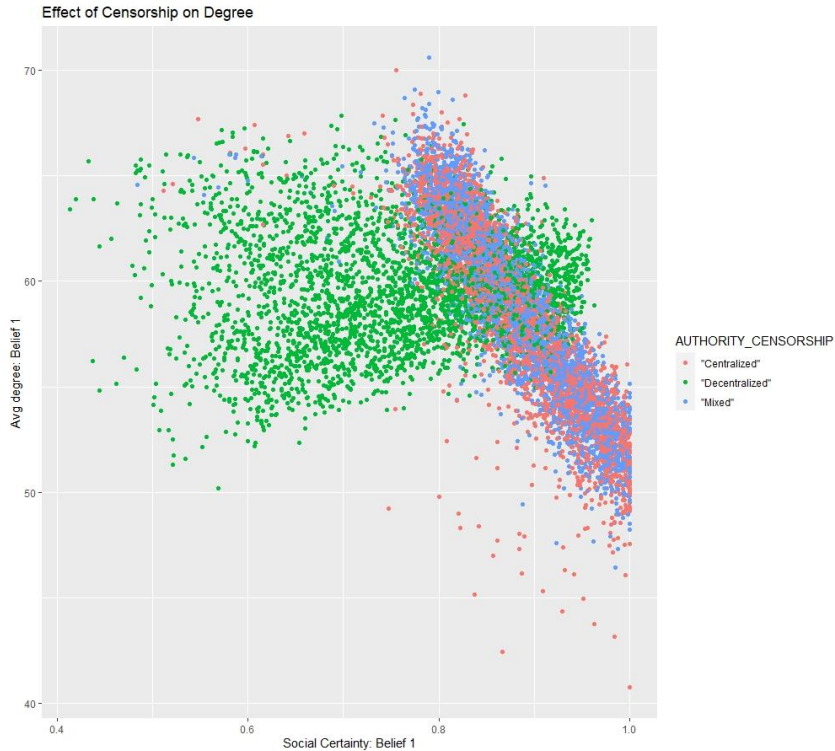

Figure 3. Effects of censorship on radicalized groups.

To further investigate this effect, we plotted the general social certainty of each group by censorship style. We found that, generally, there is little effect of censorship style on the certainty with which beliefs are held. This is true for both radical and non-radical beliefs (shown in Figures 4 and 5 below).

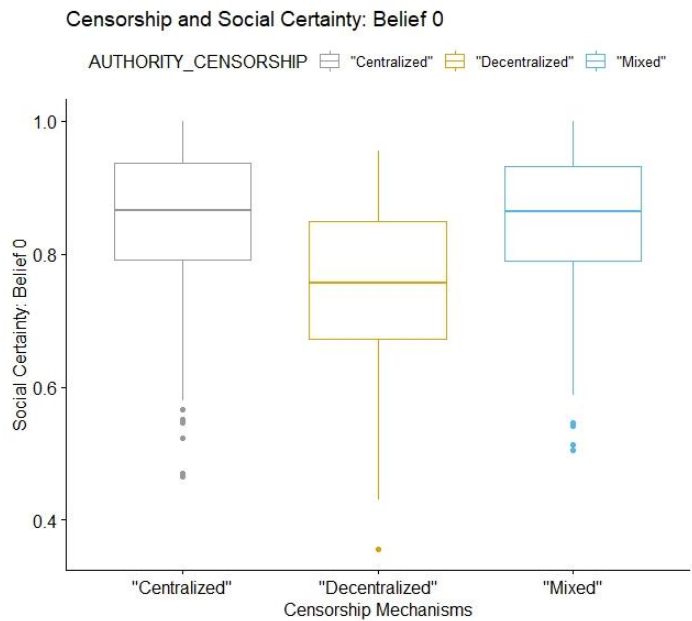

Figure 4.

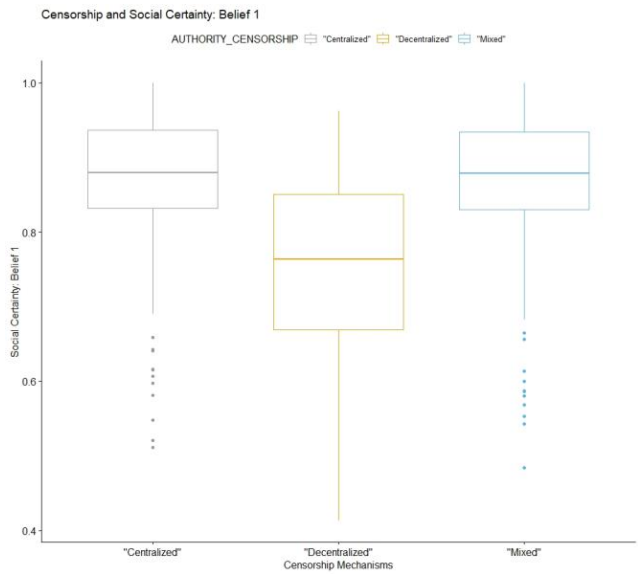

Figure 5.

A kruskal wallis test revealed that there were significant differences between the censorship groups for both belief 0 and 1. In both cases, post-hoc analysis revealed that groups under the decentralized censorship condition had significantly less belief certainty compared to the centralized and mixed groups, and there were no significant differences between groups under the centralized and mixed censorship conditions.

Additionally, in non-radicalized groups, centralized and mixed censorship created the least amount of dissent.

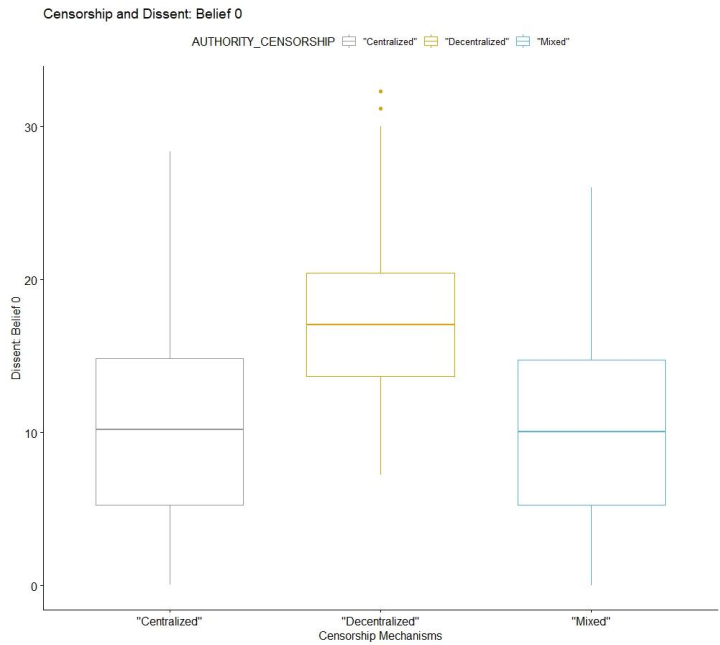

Figure 6.

However, in radicalized groups, the pattern stayed the same. While in all conditions, dissent was lower than in non-radicalized groups, when censorship was centralized or mixed, dissent was very low, while in decentralized censorship, dissent was higher.

Kruskal wallis tests also revealed that for both Belief "0" and "1" the differences in censorship mechanisms were significant. However, a post hoc anlaysis revealed that for Belief "0," the relationship between mixed and centralized was not significantly different, only their relationships with decentralized censorship. Regarding Belief "1," it was found that all three mechanisms were significantly different, with decentralized censorship producing greater dissent, followed by centralized, followed by mixed (Figure 7). Notice that decentralized censorship mechanisms lead to what might be regarded as a healthy amount of dissent, while centralized and mixed censorship mechanisms lead to almost zero dissent – a result that would lower cognitive dissonance but also, potentially, reduce cognitive flexibility because individuals would be less likely to encounter information that could challenge their radicalized beliefs.

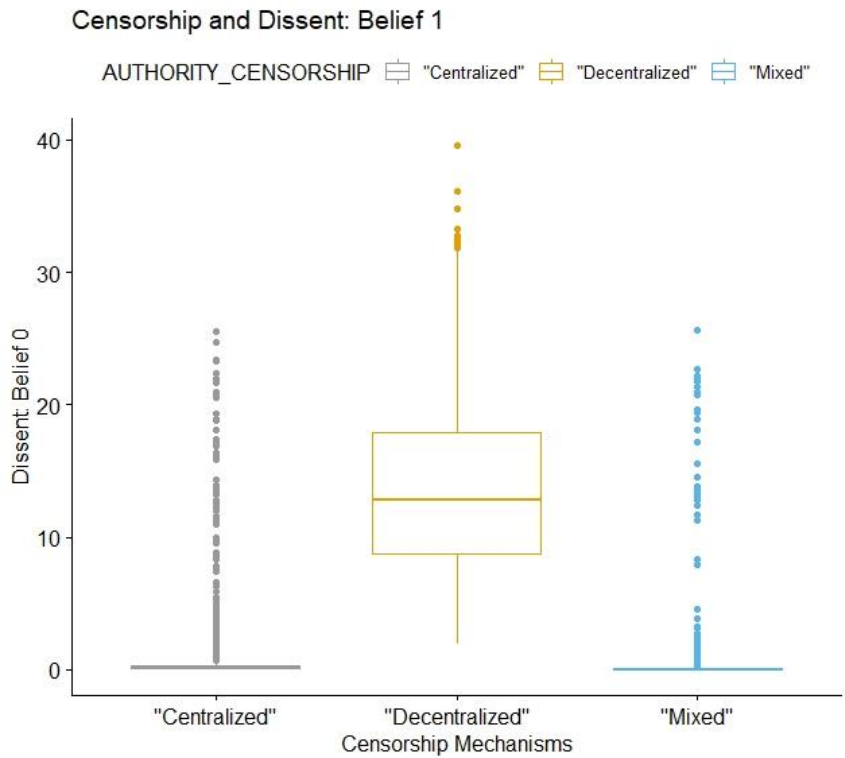

Figure 7.

The relationship between average degree and certainty in belief to level of dissent, which is one of the key findings of the model, is more clearly portrayed in Figure 8 (Belief 0) and Figure 9 (Belief 1).

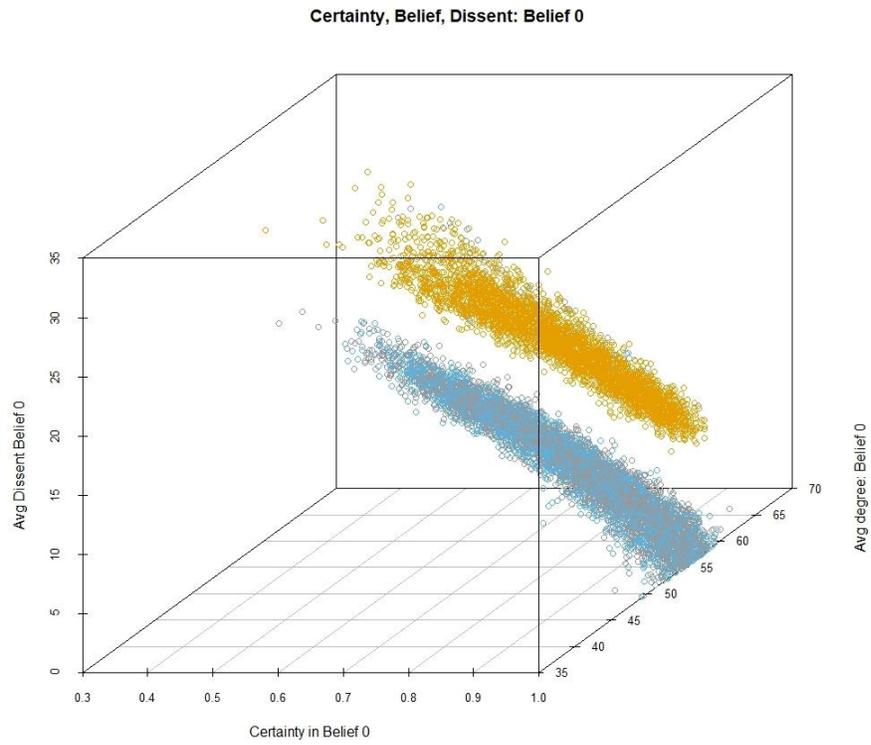

Figure 8.

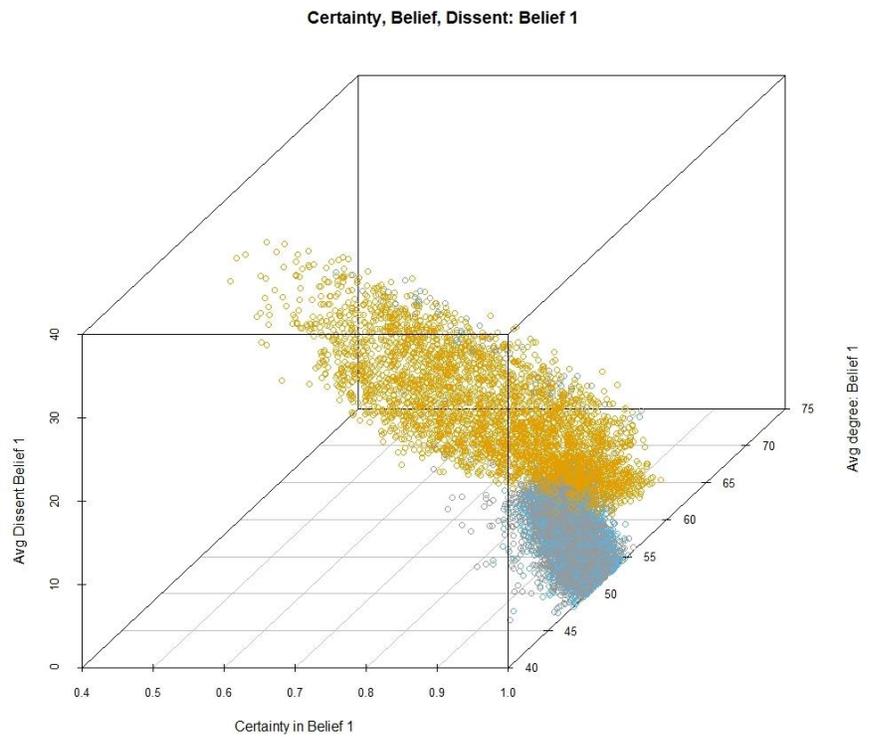

Figure 9.

These results suggest that the most efficient way to combat radicalized views in online social media populations is through decentralized censorship mechanisms, or by allowing individual users to pick and choose with whom they want to interact, rather than through centralized censorship mechanisms such as banning individuals from platforms.

## Conclusion

The results of our agent-based model and associated simulation experiments suggest that centralized censorship mechanisms can exacerbate radicalization in online social networks. Banning users for holding (or expressing) a radical belief can motivate such individuals to seek others with similar beliefs which can, in turn, lead to the creation of new, relatively isolated, online communities of substantial size. Moreover, our results suggest that members of the radicalized community may become more certain in their beliefs because they encounter less dissent, even though they have similar degrees of social support. These findings suggest that radicalized views of online communities are reinforced far more than they would be if users were allowed to curate or select, for themselves, who they want to be in their social network.

Translating these results into real world applications and risks, it suggests that the over-policing of online social networks for dissenting viewpoints might actually be counterproductive in the fight against extremism. While it may serve to create "safer" spaces for online communities, the overall health of the real-world community could be viewed as compromised when extremist views are not dealt with, but digitally "swept under the carpet". As the rise in lone wolf terrorism is now the overwhelming majority of terrorist attacks, it may be worth pause to note that those groups that are increasing in their lone wolf tactics are those which are typically the focus of social media bans [48], pushing those individuals into social media (and likely real-world) social isolation, and thus fitting one of the two profiles discerned in counter-terrorism literature [49].

Our research has several limitations. Though our agent-based model is well-specified, it is entirely possible that aspects of peoples' real-life social psychology may be inadequately modeled. Relatedly, we were not able here to directly assess the relative impact of the six mechanisms described above which might increase radicalization, in particular, identity uncertainty, differential association, shared negative affect/ identity fusion, group polarization, mutual verification contexts and internalized labeling. While prior work provides a sufficient basis for inferring the relevant impact of each on radicalization, we were not able here to adjudicate between mechanisms that are more (or less) influential in online radicalization. Future work should explore the respective impact of these mechanisms as well as other social psychological specifications of agents within simulations. In addition, further research using other methodologies should be carried out to test the plausibility of our hypothesis. Though we employed agent-based simulations here, future work might employ survey-based or qualitative methods to investigate subsets of banned real-world individuals in order to determine their perception of the social experience of censorship.

Despite these limitations, our work here contributes in a novel way to emerging research on online censorship by not only simulating online activity amongst individuals holding varied beliefs, but also by investigating the consequences of three different forms of online censorship. The initial findings of our agent-based model indicate that centralized censorship of individuals with radical beliefs, such as banning those who participated in and instigated the attack on the US Capitol on 6 January 2021 from Twitter, Facebook, and other social media platforms may actually accelerate, instead of abate, ideological extremism. While offensive speech and false beliefs are undoubtedly non-optimal forms of communication and information, methods of responding to these acts and beliefs may cause more harm than good, despite the best of intentions.